\documentclass[twocolumn,showpacs,preprintnumbers,amsmath,amssymb]{revtex4}

\usepackage{graphicx}
\usepackage{dcolumn}
\usepackage{bm}

\begin{document}
\bibliographystyle{apsrev}

\title{Observation of strong electron dephasing in disordered Cu$_{93}$Ge$_4$Au$_3$ thin films}

\author{S. M. Huang,$^{1,2}$ T. C. Lee,$^3$ H. Akimoto,$^4$ K. Kono,$^2$ and J. J. Lin,$^{1,3,\ast}$}
\affiliation{$^1$Institute of Physics, National Chiao Tung University, Hsinchu 30010, Taiwan\\
$^2$Low Temperature Physics Laboratory, RIKEN, Hirosawa 2-1, Wako-shi, Saitama 351-0198, Japan\\
$^3$Department of Electrophysics, National Chiao Tung University, Hsinchu 30010, Taiwan\\
$^4$Nano-Science Joint Laboratory, RIKEN, Hirosawa 2-1, Wako-shi, Saitama 351-0198, Japan}

\date{\today}

\begin{abstract}

We report the observation of strong electron dephasing in a series of disordered
Cu$_{93}$Ge$_4$Au$_3$ thin films. A very short electron dephasing time possessing very weak
temperature dependence around 6 K, followed by an upturn with further decrease in temperature
below 4 K, is found. The upturn is progressively more pronounced in more disordered samples.
Moreover, a ln$T$ dependent, but high-magnetic-field-insensitive, resistance rise persisting from
above 10 K down to 30 mK is observed in the films. These results suggest a nonmagnetic dephasing
process which is stronger than any known mechanism and may originate from the coupling of
conduction electrons to dynamic defects.

\end{abstract}
\maketitle


The electron dephasing time, $\tau_\varphi$, is the key quantity in the mesoscopic physics.
Recently, intense theoretical \cite{Zaikin98,Zawa99,Imry99,Galperin04,Imry03} and experimental
\cite{Mohanty97,Natelson01,Pierre03,Bauerle03} efforts have been made to address the low
temperature behavior of $\tau_\varphi$. In particular, the issue of whether the value of
$\tau_\varphi$ should saturate or diverge as $T \rightarrow 0$ K is being discussed. A
``saturated" $\tau_\varphi$ would imply a breakdown of the Fermi-liquid picture in a mesoscopic
system at very low $T$ \cite{Mohanty97}. The inelastic scattering between the conduction electrons
and a localized moment in a metal host containing dilute magnetic impurities has recently been
intensely studied in this context \cite{Zarand04,Micklitz06,Kettemann,Bauerle03,Mallet,Alzoubi}.
It has also been argued if intrinsic electron processes might lead to a saturation in
$\tau_\varphi (T \rightarrow 0)$ \cite{Zaikin98}. In addition, there are theories suggesting that
materials properties associated with dynamical structural defects may cause profound dephasing at
low $T$ \cite{Zawa99,Imry99,Galperin04,Imry03}. Despite these different proposals, it is accepted
that the responsible electron dephasing processes in ({\em comparatively}) {\em highly} and {\em
weakly disordered} metals are {\em dissimilar} \cite{Pierre03,Zawa05,Imry03}. That is, while it is
realized that magnetic scattering is important in weakly disordered metals \cite{Zarand04}, a
different mechanism is believed to be relevant for the weak $T$ dependence of $\tau_\varphi$ found
in more disordered alloys \cite{Ovadyahu83,Ovadyahu84,Ovadyahu01,Lin02EPL,Lin01JPCM}. Thus, to
understand the issue of the interactions between conduction electrons and low-lying excitations,
and, more importantly, to resolve the low-$T$ dephasing problem, information about $\tau_\varphi$
in highly disordered metals is indispensable. In this work, we have measured $\tau_\varphi$ as a
function of $T$ and the sheet resistance, $R_\Box$, in a series of two-dimensional (2D)
Cu$_{93}$Ge$_4$Au$_3$ thin films. Our values of the electron diffusion constant $D$ ($\propto
\rho^{-1}$, $\rho$ is the resistivity) are a factor $\sim$ several tens smaller than those ($D
\approx$ 100$-$200 cm$^2$/s) in the metal wires recently studied in Refs.
\cite{Mohanty97,Pierre03,Alzoubi,Mallet}. Our results provide strong evidence that the origin for
the saturation in $\tau_\varphi$ is {\em nonmagnetic}.

\begin{figure}
\includegraphics[scale=0.65]{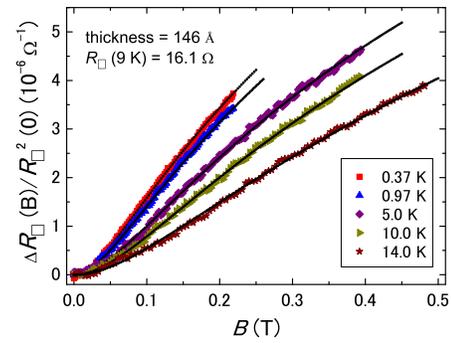}
\caption{\label{f1} (color online) Perpendicular MRs for a Cu$_{93}$Ge$_4$Au$_3$ thin film at
several temperatures as indicated. The dotted curves are the 2D WL theoretical predictions.}
\end{figure}

We have chosen Cu$_{93}$Ge$_4$Au$_3$ (CuGeAu for short) as our system material. Ge atoms were
doped into the Cu host mainly to increase the impurity scattering to enhance the weak-localization
(WL) effects, while Au atoms were doped mainly to introduce spin-orbit scattering. In the limit of
strong spin-orbit scattering, $\tau_\varphi$ becomes the {\em only} free parameter in the
comparison of the experimental magnetoresistances (MRs) with the WL theory \cite{Hikami80}, making
the extraction of $\tau_\varphi$ highly reliable. (Good fits of the 2D WL theories to our measured
perpendicular MRs were obtained for all films, and are shown for a representative film in
Fig.~\ref{f1}.) The starting CuGeAu target was intentionally chosen to be only of a ``medium"
purity (99.99\%) while, on the other hand, the spectrographic analysis of the target indicated low
levels of 4 (Fe), 0.3 (Mn) and 0.003 (Cr) ppm of magnetic impurities.

Our films were made by dc sputtering deposition on glass substrates held at ambient temperature.
The deposition rate was adjusted to tune the level of disorder of the films. The films were of
meander shape, defined by either a mechanical mask or photolithography, with typical length of
$\gtrsim$ 5.7 mm, width of $\sim$ 0.1 mm, and thickness of either $\approx$ 147$\pm$3 or 195$\pm$5
$\rm\AA$. Our values of $R_\Box$(9\,K) varied from 9.14 to 62.9 $\Omega$, corresponding to $D
\approx$ 4$-$25 cm$^2$/s and $k_Fl \approx$ 10$-$65 ($k_F$ is the Fermi wave number, and $l$ is
the electron mean free path). Here $D = v_Fl/3$ was evaluated using the Fermi velocity $v_F ({\rm
Cu}) = 1.57 \times 10^6$ m/s, and $l$ was computed using $\rho l = 6.4 \times 10^{-8}$ $\mu
\Omega$\,cm for Cu. The resistance ratio $R_\Box$(300\,K)/$R_\Box$(9\,K) = 1.1042$-$1.1109. For
complementariness, we have also studied a few 3D ($\gtrsim$ 0.5 $\mu$m thick) films made from the
{\em same} target, whose resistivity ratio $\rho$(300\,K)/$\rho$(10\,K) = 1.047$-$1.070. Such low
values of residual resistance ratios reflect that our sputtered samples must contain large amounts
of (structural) defects. To avoid electron overheating, the condition for equilibrium $eV_{\cal E}
\ll k_BT$ was assured in all resistance and MR measurements, where $V_{\cal E}$ is the applied
voltage across the energy relaxation length \cite{Ovadyahu01}.


\begin{figure}
\includegraphics[scale=0.72]{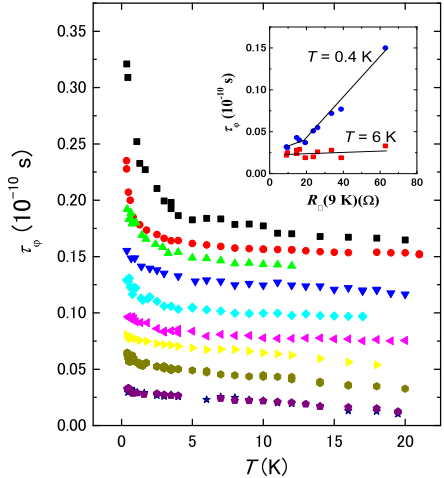}
\caption{\label{f2} (color online) $\tau_\varphi$ as a function of $T$ for CuGeAu thin films with
$R_\Box$(9\,K) (from bottom up) = 9.14, 9.66, 14.7, 16.1, 19.3, 23.7, 26.0, 33.6, 38.8, and 62.9
$\Omega$. Except for the lowest two curves, each other curve has been consecutively shifted up by
a value of 0.002 ns from its lower-neighbor curve. The inset shows the variation of the measured
$\tau_\varphi$(0.4\,K) and $\tau_\varphi$(6\,K) with $R_\Box$(9\,K). The solid lines are guides to
the eye.}
\end{figure}

Figure \ref{f2} shows the main result of this work, i.e., the measured $\tau_\varphi$ as a
function of $T$ for a series of thin films with different values of $R_\Box$. In Fig. \ref{f2},
the values of $\tau_\varphi$ for each film have been vertically shifted for clarity (notice that
$R_\Box$ increases progressively from the bottom to the top). One of the most distinct features
revealed in Fig. \ref{f2} is that $\tau_\varphi$ possesses a {\em very weak $T$ dependence} around
6 K, being {\em very short}, and has a {\em very similar magnitude} in this ``plateau" regime. The
inset indicates an almost constant $\tau_\varphi$(6\,K) $\approx$ 0.002$-$0.003 ns for {\em all}
films. The second distinct feature is that there is an upturn in $\tau_\varphi$ with decreasing
$T$ below about 4 K. In particular, the upturn is sample dependent, being {\em progressively
stronger} in {\em dirtier} films. The inset shows our measured values of $\tau_\varphi$(0.4\,K)
and $\tau_\varphi$(6\,K) versus $R_\Box$(9\,K). It depicts that $\tau_\varphi$(0.4\,K)
monotonically increases with increasing disorder. Such upturn behavior is not seen in those recent
measurements on weakly disordered samples \cite{Mohanty97,Pierre03,Alzoubi,Mallet}. Thirdly, the
electron-phonon ($e$-p) scattering rate, $\tau_{ep}^{-1} \propto T^2$ \cite{Lin02R}, is found to
be important only at $T \gtrsim$ 10 K. On the other hand, we note that the 2D electron-electron
($e$-$e$) scattering rate, $\tau_{ee}^{-1}$, which is often the dominant dephasing process in thin
films at low $T$ \cite{Alt87R}, is {\em 2 to 3 orders of magnitude smaller} than the measured
$\tau_\varphi^{-1}$ shown in Fig. \ref{f2}, and thus can be {\em ignored} in the following
analysis \cite{2Dee}.

As a first check of the $T$ dependence of our measured $\tau_\varphi$ below 5 K, we write an
effective power law $\tau_\varphi^{-1} \propto T^p$ and compare it with our data to extract the
value of $p$. We found that, even in the two most disordered films [$R_\Box$(9\,K) = 38.8 and 62.9
$\Omega$] where the upturn is most profound, the value of $p$ is still small ($\approx$
0.57$\pm$0.06). In other less disordered films, $p$ is even much smaller, being close to zero in
our cleanest films. In all cases, our values of $p$ are {\em markedly lower} than that ($p$ = 1)
recently predicted in the theory for soft local defects induced dephasing at temperatures below
the ``plateau" regime \cite{Imry03}.

Concerning the observation of a very weak $T$ dependent $\tau_\varphi$, one immediately suspects
if such behavior might be due to spin-spin scattering in the presence of dilute magnetic
impurities. As a quick check, we notice that Cu could possibly contain trace Cr, Mn, or Fe
impurities and form Kondo alloys. In a Kondo system, the magnetic scattering rate, $\tau_m^{-1}$,
is maximum at $T = T_K$, where also a ``plateau" in the dephasing time, $\tau_\varphi^{\rm
plateau}$, often appears. Based on the Nagaoka-Suhl (NS) expression \cite{Haesendonck87}, Pierre
{\it et al.} \cite{Pierre03} pointed out that for Cu, $\tau_\varphi^{\rm plateau} = (c_m/\pi \hbar
\nu)^{-1}$ $\simeq$ (0.6/$c_m$) ns, where $c_m$ is the magnetic impurity concentration in ppm, and
$\nu$ is the total electron density of states at the Fermi level. Then, our experimental value of
$\tau_\varphi$(6\,K) $\approx$ 0.002$-$0.003 ns would imply a level of $c_m \approx$ 200$-$300 ppm
in the samples, if our measured $\tau_\varphi^{-1}$ were directly ascribed to $\tau_m^{-1}$. Such
a level of $c_m$ is obviously too high to be realistic.

\begin{figure}
\includegraphics[scale=0.48]{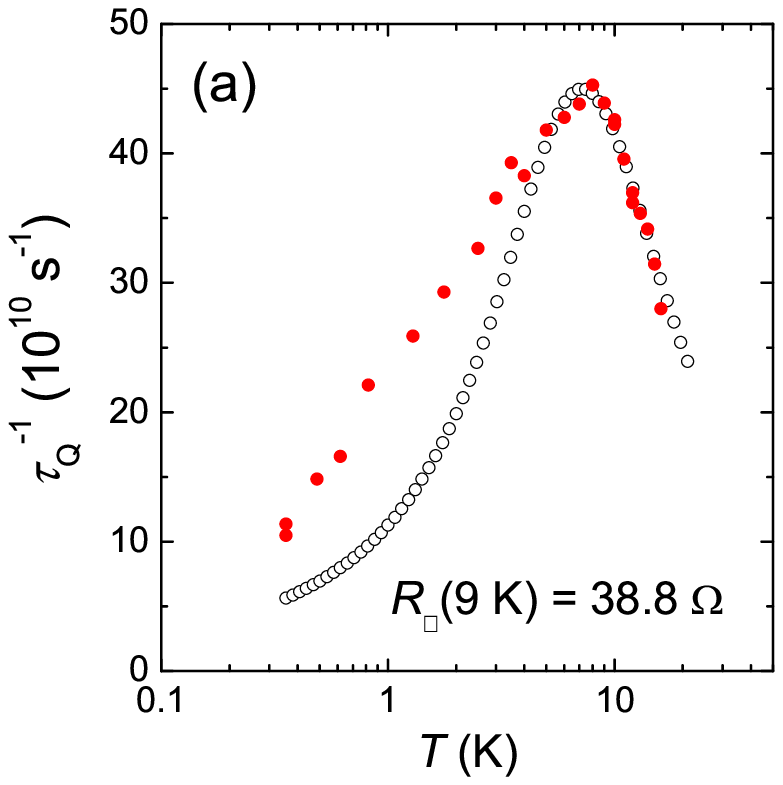}
\includegraphics[scale=0.48]{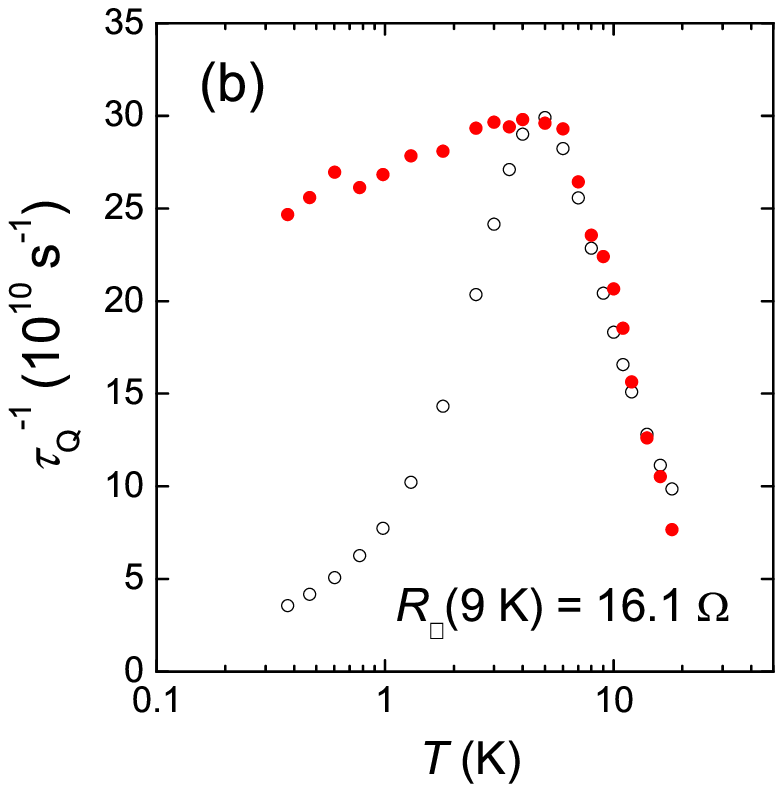}
\caption{\label{f3} (color online) $\tau_Q^{-1}$ as a function of $T$ for two CuGeAu thin films.
The solid symbols are the experimental data and the open symbols are the Nagaoka-Suhl expression,
see text.}
\end{figure}

Recently, new theories have been established and it is realized that the NS result for
$\tau_m^{-1}$ need be revised. The new calculations of Zarand {\it et al.} \cite{Zarand04} and
Micklitz {\it et al.} \cite{Micklitz06} for spin $S = \frac12$ impurities have confirmed that the
scattering rate $\tau_m^{-1}$ is maximum at $T = T_K$. Their corrected peak scattering rate is
$\approx$ 8\% lower than the NS prediction. So the estimate of $c_m$ given above will not be
substantially altered even if one applies the new theory. In fact, for all $T$, the
Zarand-Micklitz (ZM) theory predicts a $\tau_m^{-1}$ {\em below} the corresponding NS scattering
rate. The later is in turn {\em below} our experimental scattering rate (at low temperatures, see
Fig. \ref{f3}).

To unravel the intriguing dephasing mechanism responsible for our measured $\tau_\varphi$, we
carry out further quantitative analysis below. Assuming that our measured $\tau_\varphi^{-1}$
between 0.3 and 25 K is given by $\tau_\varphi^{-1} = A_{ep}T^2 + \tau_Q^{-1}$ (recall that
$\tau_{ee}^{-1}$ is totally negligible), where $\tau_Q^{-1}$ denotes a yet-to-be identified
dephasing rate, and the $e$-p coupling strength $A_{ep}$ can be determined from the high-$T$ part
of the measured $\tau_\varphi$ \cite{Lin02R}. In Fig. \ref{f3}, we plot the variation of
$\tau_Q^{-1}$ with $T$ for two representative films. Figure \ref{f3} clearly reveals a maximum in
$\tau_Q^{-1}$ at a characteristic temperature which we denote as $T_K^\ast$. (We will see later
that this $T_K^\ast$ cannot be identified with the Kondo temperature $T_K$.) Below $T_K^\ast$, one
sees that the $T$ dependence of $\tau_Q^{-1}$ is much stronger for the higher $R_\Box$ film, as
can be expected from the above discussion that the exponent $p$ decreases with decreasing
$R_\Box$.

Since at $T \gtrsim T_K$, the ZM theory \cite{Zarand04,Micklitz06} essentially reproduces the
conventional wisdom, we first compare our experimental $\tau_Q^{-1}$ with the NS expression
\cite{Haesendonck87,Bauerle05} for the temperature range $T \gtrsim T_K^\ast$. In plotting the NS
approximations in Fig. \ref{f3}, we have adjusted the free parameters ($T_K^\ast$, $c_m$, and the
local spin $S$) so that the theory reproduced the experiment at $T \gtrsim T_K^\ast$. Figure
\ref{f3} illustrates that the NS expression can describe our measurements down to slightly below
$T_K^\ast$. However, inspection of the fitted values indicates that such agreement is spurious,
because such good fits can only be achieved by using {\em unrealistic} values for the adjusting
parameters. For example, a local spin of $S$ = 0.12 (0.082) and $T_K^\ast$ = 7.2 (4.75) K had to
be used for the $R_\Box$(9\,K) = 38.3 (16.1) $\Omega$ film. Using $S = \frac12$ or any larger
value can {\em never} reproduce our data. Moreover, if we ascribe the measured
($\tau_Q^{-1}$)$^{max}$ to $\tau_m^{-1}(T = T_K)$, an unreasonably large value of $c_m$ =
200$-$300 ppm will be inferred, implying that an {\em unusually strong} $\tau_Q^{-1}$ is entirely
dominating over the $e$-p and $e$-$e$ scattering in our films in this $T$ range. (If we apply the
ZM theory \cite{Zarand04,Micklitz06}, the estimate for $c_m$ would be even higher by $\approx$
8\%.) Although accidentally formed CuO on the film surfaces may have $S$ = 1 spin
\cite{Vranken88}, one would not expect a huge $c_m$ $>$ 200 ppm to result from such oxidation.
Besides, there is {\em no} known Cu based Kondo alloys which have values of $T_K$ around $\approx$
5$-$7 K. Thus, in all aspects, it is certainly impossible to ascribed our measured $\tau_Q^{-1}$
to magnetic scattering.

\begin{figure}
\includegraphics[scale=0.6]{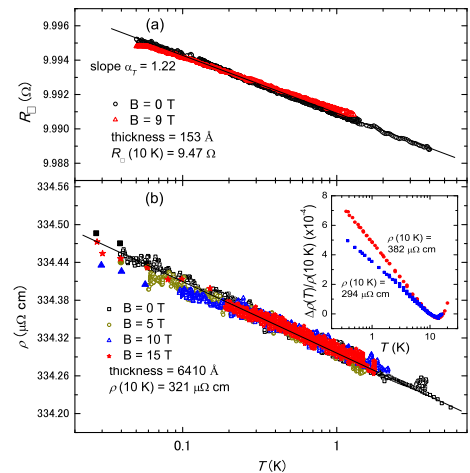}
\caption{\label{f4} (color online) (a) Variation of $R_\Box$ with $T$ for a thin film. (b)
Variation of $\rho$ with $T$ for a thick film. Inset: $\triangle \rho(T)/\rho(10\,{\rm K}) =
[\rho(T) - \rho(10\,{\rm K}] / \rho(10\,{\rm K})$ as a function of $T$ for two thick films.}
\end{figure}

Another decisive way of checking whether magnetic impurities might exist to a degree in the
samples is to explore if the variation of resistance with $T$ reveals any Kondo or spin-glass
behavior. Figure \ref{f4}(a) shows the variation of $R_\Box$ with $T$ for a representative thin
film in zero field and in a high perpendicular magnetic field $B$. This figure clearly indicates
that, both in $B$ = 0 and 9 T, the $R_\Box$ reveals a ln$T$ behavior all the way down to 50 mK.
There is not even a sign of a crossover to a saturation characterizing a Kondo system in the
unitary limit \cite{Mallet,Daybell67}. In fact, if there were $c_m \gtrsim$ 200 ppm in the films,
we should have been in a spin-glass state at low $T < T_K$ and a marked decrease in $R_\Box$
should be observed as the local moments freeze into a collective state \cite{Monod67}. Moreover,
in the presence of a large magnetic field $B \gg k_BT/g\mu_B$ ($\mu_B$ is the Bohr magneton), the
local spins should be aligned and a notable decrease in $R_\Box$ should occur at low $T$
\cite{Monod67}. However, {\em none} of these features are found in Fig. \ref{f4}(a). Thus, our
$R_\Box$ as a function of $T$ and $B$ does {\em not} support a picture based on magnetic
impurities.

Usually, a ln$T$ increase in $R_\Box$ in thin metal films may be due to WL and $e$-$e$ interaction
(EEI) effects, and the resistance rise can be written as $\triangle R_\Box (T)/R_\Box^2 = -
\alpha_T (e^2/2 \pi^2 \hbar)\, {\rm ln} T$, where $\alpha_T = \alpha p + (1 - F^\prime)$. The
parameter $\alpha p$ characterizes the WL effect and $1-F^\prime$ the EEI effect. For disordered
metal films in the limit of strong spin-orbit interaction, the screening factor $F^\prime$ is
small (typically, $\lesssim$ 0.1) and the EEI effect dominates the resistance rise while the WL
contribution is negligibly small \cite{Alt87R}. Then, these two effects would cause a maximum
resistance rise with a slope $\alpha_T = 1$. For all the thin films studied in this work, we
obtain $\alpha_T = 1.33 \pm 0.13$. This value is systematically larger than unity, strongly
implying that there must be an extra mechanism which also contributes to the ln$T$ rise in
$R_\Box$. This additional contribution is, however, insensitive to high magnetic fields.

To further illuminate this anomalous ln$T$ resistance rise, we have also made a few {\em thick}
films from the {\em same} sputtering target under {\em similar} deposition conditions and measured
the resistance. Figure \ref{f4}(b) shows the variation of $\rho$ with $T$ for a representative
thick film in zero field and in perpendicular magnetic fields. Usually, for a disordered bulk
metal, one expects to see a resistance rise obeying a $- \sqrt{T}$ law \cite{Alt87R}. This figure,
however, reveals a strict ln$T$ behavior down to 30 mK, indicating some mechanism being dominating
over the 3D EEI effect in our samples. (The inset demonstrates that the ln$T$ law is robust,
persisting up to $>$ 10 K.) Moreover, there is definitely no sign of a saturated (decreased)
resistance signifying the presence of the Kondo (spin-glass) behavior. Also similar to the case of
thin films, there is no evidence of profound negative MRs in the presence of a high $B$. These
results strongly suggest that magnetic scattering, if any exists, could not be the primary
mechanism in our samples. On the other hand, it has recently been reported that scattering of
electrons off two-level systems can cause a ln$T$ and high-$B$-insensitive variation of the
resistance down to very low temperatures \cite{Cichorek03}. Our results in Fig. \ref{f4} are in
line with this observation. If dynamic defects are somehow abound in our sputtered films, then it
is natural to ask to what extent such defects can contribute to dephasing
\cite{Zawa99,Imry99,Zawa05,dynamic}.

Theoretically, for {\em highly disordered} 3D systems, a $\tau_\varphi$ possessing a very weak $T$
dependence in a certain temperature interval and then crossing over to a {\em slow increase} with
{\em decreasing} $T$ has recently been predicted in a model based on tunneling states of dynamical
structural defects \cite{Galperin04}. This model also predicts a ``counterintuitive" scaling
$\tau_\varphi \propto \rho$ in the plateau-like region. Our observations in Fig. \ref{f2}
essentially mimic these {\em qualitative} features. How this theory may be generalized to the 2D
case should be of great interest. Experimentally, a scaling $\tau_\varphi \propto D^{-1}$
($\propto \rho$) for a good number of 3D {\em polycrystalline} alloys has recently been reported
by Lin and Kao \cite{Lin01JPCM}. A dephasing time $\tau_\varphi (\text{10\,K})$ increasing with
disorder $(k_Fl)^{-1}$ has been reported for 3D In$_2$O$_{3-x}$ thick films \cite{Ovadyahu84}, and
a $\tau_\varphi (\text{4.2\,K}) \propto \rho$ has been reported for 2D In$_2$O$_{3-x}$ thin films
\cite{Ovadyahu83} by Ovadyahu. Those results and Fig.~\ref{f2} may indicate some generality of the
dynamic defects in causing dephasing at low $T$ \cite{Galperin04,Imry03}. This, yet-to-be fully
understood, {\em nonmagnetic} dephasing mechanism deserves further theoretical and experimental
investigations. It should also be reemphasized that the {\em anomalous} dephasing time found in
the present work is 2 to 3 orders of magnitude {\em shorter} than the 2D $e$-$e$ scattering time
\cite{Alt87R}.


We report the observation of a strong electron dephasing in disordered Cu$_{93}$Ge$_4$Au$_3$ thin
films. This dephasing is much stronger than any known inelastic electron scattering process. Our
observation of a strict ln$T$ dependent, but high-magnetic-field-insensitive, resistance rise
indicates that this dephasing must be nonmagnetic in origin. The recent theoretical concept
\cite{Galperin04,Imry03} of the dynamic defects is qualitatively in line with our result. This
work provides key information for uncovering the long-standing saturation problem of
$\tau_\varphi$ in mesoscopic systems.


We acknowledge helpful discussions with C. B\"auerle, Y. M. Galperin, V. Vinokur, A. Zaikin, G.
Zarand, and A. Zawadowski. This work was supported by the Taiwan NSC through Grant No. NSC
94-2112-M-009-035, by the MOE ATU Program, and by the RIKEN-NCTU Joint Graduate School Program (to
SMH).


\begin{references}

\item[$^\ast$] Electronic address: jjlin@mail.nctu.edu.tw

\bibitem{Zaikin98} D.S. Golubev and A.D. Zaikin, Phys. Rev. Lett. \textbf{81}, 1074 (1998).

\bibitem{Zawa99} A. Zawadowski {\it et al.}, Phys. Rev. Lett. \textbf{83}, 2632 (1999).

\bibitem{Imry99} Y. Imry {\it et al.}, Europhys. Lett. \textbf{47}, 608 (1999).

\bibitem{Galperin04} Y.M. Galperin {\it et al.}, Phys. Rev. B \textbf{69},
073102 (2004); V.V. Afonin {\it et al.}, {\it ibid.} \textbf{66}, 165326 (2002).

\bibitem{Imry03} Y. Imry {\it et al.}, arXiv:cond-mat/0312135.

\bibitem{Mohanty97} P. Mohanty {\it et al.}, Phys. Rev. Lett. \textbf{78},
3366 (1997); P. Mohanty and R.A. Webb, {\it ibid.} \textbf{91}, 066604 (2003).

\bibitem{Natelson01} D. Natelson {\it et al.}, Phys. Rev. Lett. \textbf{86}, 1821 (2001).

\bibitem{Pierre03} F. Pierre and N.O. Birge, Phys. Rev. Lett. {\bf 89}, 206804 (2002);
F. Pierre {\it et al.}, Phys. Rev. B \textbf{68}, 085413 (2003).

\bibitem{Bauerle03} F. Schopfer {\it et al.}, Phys. Rev. Lett. \textbf{90}, 056801 (2003).

\bibitem{Zarand04} G. Zarand {\it et al.}, Phys. Rev. Lett. \textbf{93}, 107204 (2004).

\bibitem{Micklitz06} T. Micklitz {\it et al.}, Phys. Rev. Lett.
\textbf{96}, 226601 (2006); T. Micklitz {\it et al.}, Phys. Rev. B \textbf{75}, 054406 (2007).

\bibitem{Kettemann} S. Kettemann and E. R. Mucciolo, Phys. Rev. B \textbf{75}, 184407 (2007).

\bibitem{Mallet} F. Mallet {\it et al.}, Phys. Rev. Lett. \textbf{97}, 226804 (2006).

\bibitem{Alzoubi} G.M. Alzoubi and N.O. Birge, Phys. Rev. Lett. \textbf{97}, 226803 (2006).

\bibitem{Zawa05} O. Ujsaghy and A. Zawadowski, J. Phys. Soc. Jpn. \textbf{74}, 80 (2005).

\bibitem{Ovadyahu83} Z. Ovadyahu, J. Phys. C: Solid State Phys. \textbf{16}, L845 (1983).

\bibitem{Ovadyahu84} Z. Ovadyahu, Phys. Rev. Lett. \textbf{52}, 569 (1984).

\bibitem{Ovadyahu01} Z. Ovadyahu: Phys. Rev. B \textbf{63}, 235403 (2001).

\bibitem{Lin02EPL} J.J. Lin {\it et al.}, Europhys. Lett. \textbf{57}, 872 (2002);
J.J. Lin and N. Giordano, Phys. Rev. B \textbf{35}, 1071 (1987); J.J. Lin {\it et al.}, J. Phys.
Soc. Jpn. \textbf{72}, 7 (2003), Suppl. A.

\bibitem{Lin01JPCM} J.J. Lin and L.Y. Kao, J. Phys.: Condens. Matter \textbf{13}, L119 (2001).

\bibitem{Hikami80} S. Hikami {\it et al.}, Prog. Theor. Phys. \textbf{63}, 707 (1980).

\bibitem{Lin02R} J.J. Lin and J.P. Bird, J. Phys.: Condens. Matter \textbf{14}, R501 (2002).
For our films, $\tau_{ep} \approx (0.53 \pm 0.1) T^{-2}$ ns\,K$^{-2}$.

\bibitem{Alt87R} B. L. Altshuler {\it et al.}, Sov. Sci. Rev. A \textbf{9}, 223 (1987).

\bibitem{2Dee} For a typical film with $R_\Box$ = 30 $\Omega$, the Eq. (2.36) of Ref.
\cite{Alt87R} predicts $\tau_{ee} \approx 1.1 T^{-1}$ ns\,K$^{-1}$.

\bibitem{Haesendonck87} C. Van Haesendonck {\it et al.}, Phys. Rev.
Lett. \textbf{58}, 1968 (1987); R.P. Peters {\it et al.}, {\it ibid.} \textbf{58}, 1964 (1987).


\bibitem{Bauerle05} C. B\"auerle {\it et al.}, Phys. Rev. Lett. \textbf{95}, 266805 (2005).

\bibitem{Vranken88} J. Vranken {\it et al.}, Phys. Rev. B \textbf{37}, 8502 (1988).

\bibitem{Daybell67} M.D. Daybell and W.A. Steyert, Phys. Rev. Lett. \textbf{18}, 398 (1967).

\bibitem{Monod67} P. Monod, Phys. Rev. Lett. \textbf{19}, 1113 (1967).

\bibitem{Cichorek03} T. Cichorek {\it et al.}, Phys. Rev. B \textbf{68}, 144411 (2003).

\bibitem{dynamic} The dynamic defects might be associated with the doped Au or Ge atoms. However,
the precise nature requires further studies.


\end{references}
\end{document}